\documentclass[prm,reprint,amsmath,amssymb,showpacs,floatfix]{revtex4-1}
\usepackage{graphicx}
\usepackage{natbib}
\usepackage{bm}
\usepackage{float}
\usepackage{array}
\usepackage[usenames]{color}
\usepackage[export]{adjustbox}
\graphicspath{ {./figures_SI/}}

\newcommand{\vs}{$V_{S}^{2-}$}

\begin{document}

\title{Localized Excitons in Defective Monolayer Germanium Selenide}

\author{Arielle Cohen}
\affiliation{Division of Materials Science and Engineering, Boston University, Boston, MA 02215}
\author{D. Kirk Lewis}
\affiliation{Department of Electrical and Computer Engineering, Boston University, Boston, MA 02215}
\author{Tianlun Huang}
\affiliation{Division of Materials Science and Engineering, Boston University, Boston, MA 02215, USA}
\author{Sahar Sharifzadeh}
\affiliation{Department of Electrical and Computer Engineering, Boston University, Boston, MA 02215}
\affiliation{Division of Materials Science and Engineering, Boston University, Boston, MA 02215}
\email{ssharifz@bu.edu}

\date{\today}

\begin{abstract}
Germanium Selenide (GeSe) is a van der Waals-bonded layered material with promising optoelectronic properties, which has been experimentally synthesized for 2D semiconductor applications. In the monolayer, due to reduced dimensionality and, thus, screening environment, perturbations such as the presence of defects have a significant impact on its properties. We apply density functional theory and many-body perturbation theory to understand the electronic and optical properties of GeSe containing a single selenium vacancy in the $-2$ charge state. We predict that the vacancy results in mid-gap ``trap states'' that strongly localize the electron and hole density and lead to sharp, low-energy optical absorption peaks below the predicted pristine optical gap. Analysis of the exciton wavefunction reveals that the 2D Wannier-Mott exciton of the pristine monolayer is highly localized around the defect, reducing its Bohr radius by a factor of four and producing a dipole moment along the out-of-plane axis due to the defect-induced symmetry breaking. Overall, these results suggest that the vacancy is a strong perturbation to the system, demonstrating the importance of considering defects in the context of material design.
\end{abstract}

\maketitle

\section{Introduction}
Since the isolation of single layers of graphene in 2004, there has been growing interest in identifying, synthesizing and characterizing other two-dimensional (2D) materials with unique mechanical, electronic, optical, and thermal properties. Promising candidates for 2D materials have a layered structure, with strong bonds within and weaker van der Waals bonds between layers. While algorithmic searches through materials databases suggest that many hundreds of such materials exist, only a small fraction have been studied in detail, either computationally or experimentally~\cite{Cheon2017}.

Understanding the role defects play in the optoelectronic properties of 2D materials is critical to incorporating these materials into devices. Defects occur naturally in all materials, but their impact is magnified in two dimensions. First, the reduced length scale in the out-of-plane direction increases localization of electrons and holes, which are more strongly perturbed by the presence of the defect~\cite{lin_defect_2016}. Second, the reduced dielectric screening of Coulomb forces compared to a bulk solid results in an increased perturbation associated with defects~\cite{qiu_screening_2016,tongay_defects_2013}. Thus, defects can act as traps for electrons, holes or excitons~\cite{tongay_defects_2013}, creating localized electronic states that alter absorption and emission~\cite{tonndorf_single-photon_2015}, carrier mobility~\cite{zou_open_2015}, phonon scattering~\cite{parkin_raman_2016, qin_diverse_2016}, and photocatalytic properties of the 2D material~\cite{zhang_defect-rich_2018}. Depending on the desired application, these localized states may be harmful (e.g., undesired luminescence or carrier scattering~\cite{yu_defect_2018}), or advantageous in localizing excitations for photocatalysis and quantum computing applications~\cite{wang_engineering_2017,wu_first-principles_2017,zou_open_2015}). Identifying the nature of single defects experimentally is still quite challenging~\cite{barja_identifying_2019,schuler_large_2019}, and so computational studies of defects play a critical role in materials design and discovery for high-performance devices composed of 2D materials~\cite{giantomassi_electronic_2011}.

Among 2D materials, monolayer Group IV-VI monochalcogenides such as GeSe are particularly attractive for further study because they have band gaps close to or within the visible range~\cite{xu_electronic_2017} and their electronic properties are strain tunable~\cite{hu_recent_2019}. The strain tunability derives from the structure of the layers, which exhibit anisotropy and ``hinge-like'' arrangements between atoms, similar to phosphorene (the monolayer form of black phosphorus)~\cite{gomes_phosphorene_2015,hu_recent_2019}. Monolayer and few-layer GeSe, in particular, shows promise for applications in energy, sensing and electronics such as high-performance solar cells~\cite{zhao_design_2017,antunez_tin_2011}, photocatalytic water splitting~\cite{lv_two-dimensional_2017}, and LiO\textsubscript{2} battery cathodes~\cite{ji_monolayer_2017}. In addition, GeSe is technologically appealing in that germanium and selenium are relatively abundant on Earth and less dangerous than heavier elements, which are common components of other narrow band gap monolayers~\cite{vaughn_ii_single-crystal_2010}.

There have been limited studies of defects in monolayer GeSe that suggest a significant impact on its properties. Prototype photodetectors with high photoresponsivity and external quantum efficiency have been fabricated from GeSe and GeS nanosheets in the lab~\cite{ramasamy_solution_2016,lan_synthesis_2015,mukherjee_nir_2013}. For these devices, higher concentrations of defects are posited to cause degradation of performance~\cite{ramasamy_solution_2016}. Additionally, a recent combined experimental and computational study on bulk GeSe concluded that inconsistencies in the reported absorption onset were likely the result of defects and Urbach tailing creating sub-gap absorption~\cite{murgatroyd_gese_2020}. While the precise nature of the defects responsible for these observations is not known, limited computational studies based on density functional theory (DFT) have predicted the formation and transition energies associated with point defects. DFT calculations have predicted that the Se vacancy in a neutral charge state can result in mid-gap states, consistent with the lowered onset of absorption observed experimentally~\cite{gomes_vacancies_2016}. Additionally, for orthorhombic monolayer GeSe, DFT-predicted formation energies suggest that the Se vacancy is preferred to a Ge vacancy~\cite{mao_toxic_2018}, while in the buckled hexagonal structure, DFT-based studies predict that Ge is the preferred vacancy~\cite{ersan_effect_2017}.

In this article, we study the impact of the selenium vacancy in the $-2$ charge state (\vs{}) on the electronic and optical properties of orthorhombic monolayer GeSe (Figure \ref{fig:fig1}). By applying many-body perturbation theory (MBPT) within the GW approximation~\cite{hybertsen_first-principles_1985} and Bethe-Salpeter equation (BSE) approach~\cite{rohlfing_electron-hole_2000,onida_electronic_2002,deslippe_berkeleygw:_2012}, which has been shown to accurately predict the bandstructure and optical absorption spectra of solids~\cite{hedin_new_1965,hedin_effects_1970,rohlfing_electron-hole_2000,onida_electronic_2002}, we provide quantitative predictions of the defect energetics. The GW/BSE approach has been previously applied to a limited extent to understanding the excited-state properties of other defective monolayers. In particular, defect-state transition energies, spin, and excitonic properties have been characterized for hexagonal boron nitride~\cite{wang_layer_2020,smart_fundamental_2018,wu_first-principles_2017}, various transition metal dichalcogenides~\cite{schuler_large_2019,barja_identifying_2019,refaely-abramson_defect-induced_2018,naik_substrate_2018}, and phosphorene\cite{huang_structural_2016,wang_charged_2017}. Here, we focus on the localization of excitations in defective monolayer GeSe, which has not yet been studied within the GW/BSE approximation. We predict that the presence of deep mid-gap states introduced by the defect result in low-energy excitonic peaks in optical absorption. Additionally, an analysis of the electron-hole correlation function~\cite{sharifzadeh_low-energy_2013, lewis_defect-induced_2019} reveals that the two-dimensional Wannier-Mott exciton is strongly perturbed due to the presence of the defect, with localization of its envelope function and introduction of an excited-state dipole along the out-of-plane axis. 

\section{Computational Details}

DFT calculations were performed using the Quantum Espresso package~\cite{Giannozzi_2009} within the generalized gradient approximation of Perdew-Burke-Ernzerhof (PBE)~\cite{perdew_generalized_1996}. The core and nuclei of Ge and Se atoms were described by norm-conserving Goedecker-Hartwigsen-Hutter-Teter pseudopotentials~\cite{HGH_pseudo} with, respectively, 4 and 6 valance electrons treated explicitly. All geometry relaxations had maximum force and total energy criteria of better than 0.013 eV/\AA{} and $2\times10^{-8}$ eV/atom respectively, with a planewave cutoff energy of at least 2700 eV. The initial geometry of the bulk crystal was taken from the Materials Project database~\cite{Jain2013}(Structure: mp-700) and the pristine monolayer was constructed by isolating and optimizing the geometry of a single layer of the bulk, using a $16 \times 16 \times 1$ k-point grid with $16$ \AA{} of vacuum along the out-of-plane direction. The predicted in-plane lattice vectors for the monolayer were 3.98 \AA{} and 4.27 \AA{} along the a (zigzag) and b (armchair) directions, respectively. From this geometry, a $5 \times 5$  (100 atom) supercell was constructed, the defect was introduced, and the supercell was relaxed. To create the charged defect, we removed a single Se atom from the supercell and added a charge of -2 to the system. The geometry was re-optimized using a $2 \times 2 \times 1$ k-point grid shifted off of $\textbf{k} = \mathbf{0}$ ($\Gamma$) in order to reduce defect-defect interactions~\cite{lewis_defect-induced_2019}. The defect-defect separations in the final geometry were 20.0 \AA{} and 21.6 \AA{} along a and b, respectively. The final relaxed geometry of the defective structure is shown in Figure \ref{fig:fig1}. 

MBPT calculations were performed using the BerkeleyGW package~\cite{deslippe_berkeleygw:_2012} with starting DFT-PBE orbitals and energies from Quantum Espresso. The vacuum spacing between layers was reduced to $8$ \AA{} due to computational cost; in order to minimize interactions along the out-of-plane direction, a Coulomb truncation was applied. Additionally, because of the need for fine-sampling of the k-point mesh in 2D materials~\cite{qiu_screening_2016}, we used the subsampling approach of Jornada, et al.~\cite{da_jornada_nonuniform_2017} to extrapolate the dielectric function from a coarse to fine k-point grid at a lowered computational cost. GW calculations were performed using nonuniform neck subsampling (NNS) with 10 radially subsampled q-points on a regular $10 \times 10 \times 1$ grid for the pristine and $3 \times 3 \times 1$ grid for the with-defect monolayer. The dielectric function cutoff was 8 Ry, and the number of unoccupied states was 290 for pristine and 4752 for the with-defect supercell, corresponding to 70 eV above vacuum in both cases. BSE calculations were performed using clustered sampling interpolation (CSI)  with a coarse k-point grid of $20 \times 20 \times 1$ with six valance and and six conduction bands and interpolated to a fine grid of $60 \times 60 \times 1$ with two valance and two conduction bands for the pristine monolayer. For the with-defect supercell, the CSI coarse k-point grid was $6 \times 6 \times 1$ with 10 valance and and 10 conduction bands interpolated to a fine grid of $18 \times 18 \times 1$ with 3 valance and 3 conduction bands. To compute the electron-hole correlation function, the integrals of Equation \ref{ecfint} are calculated as discrete sums over $\textbf{r}_{h}$ and $\textbf{r}_{e}$ with 75 randomly sampled hole locations within two nearest neighbor shells around the vacancy within the a-b plane, and within $\sim 1$ \AA{} above and below the monolayer.

For charged defects, we apply a post-self-consistent-field correction to the defect-localized state energies to correct for the artificial interaction of a charged defect with its periodic images~\cite{Jain2011,Chen2013}. For the three defect-localized bands, identified based on the spatial extent of the electron density (see supplemental materials~\cite{cohen2020SM}), following ~\cite{lewis_quasiparticle_2017}, we applied a correction of 0.45 eV, calculated from the COFFEE code~\cite{Naik2018} as
\begin{equation} \label{ecorr}
    \epsilon_{d,corr} = -\frac{2}{q}E_{corr},\\
\end{equation}
where $q$ is the charge state of the defect and $E_{corr}$ is the total electrostatic energy correction proposed by Freysoldt, Neugebauer, and Van de Walle~\cite{Freysoldt2011}, calculated to be 0.45 eV for this system.

\section{Results}
\begin{figure}[htbp]
    \centering
    \includegraphics[width=1.0\linewidth]{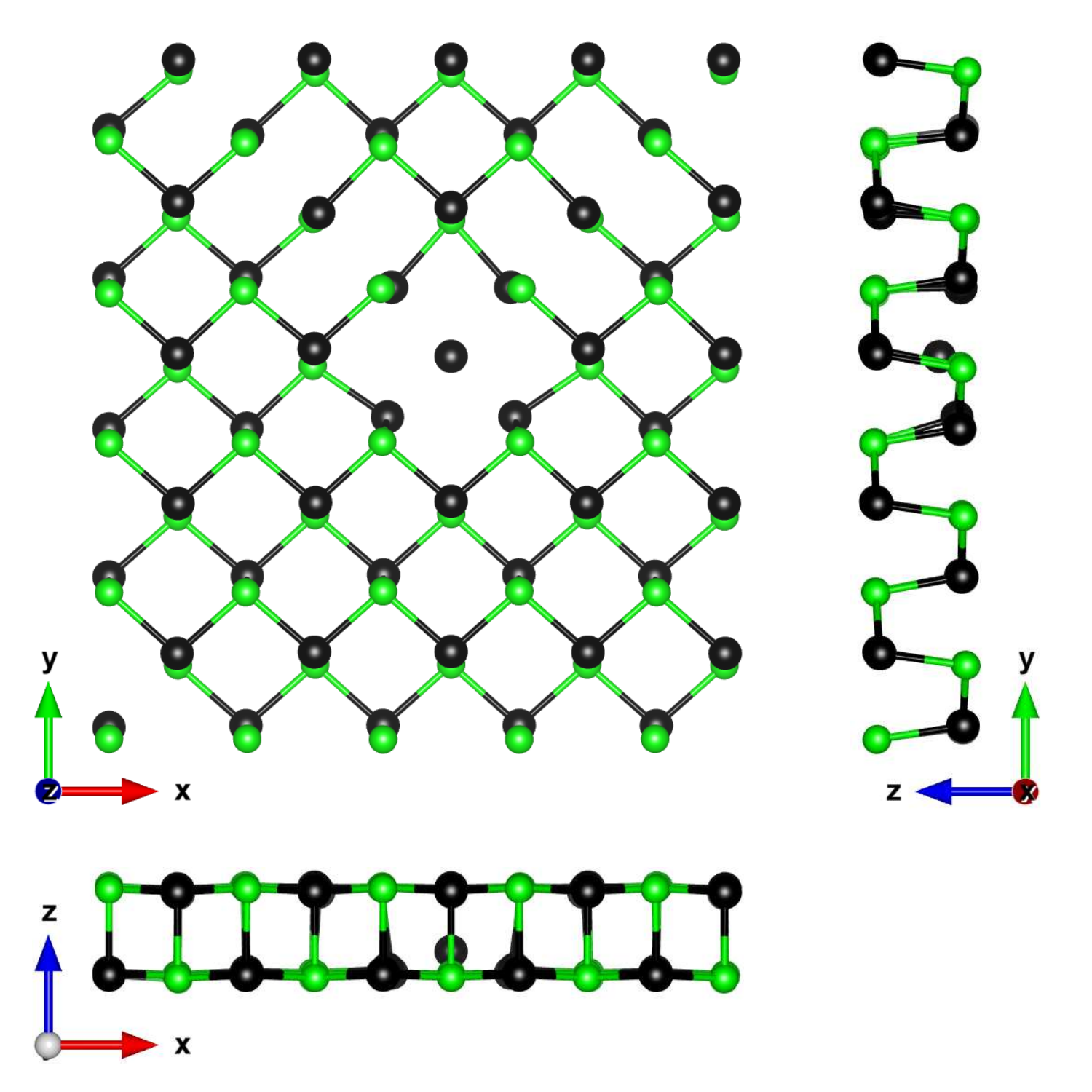}
    \caption{The structure of a 5 x 5 orthorhombic GeSe supercell containing a \vs{} point defect. Ge atoms are shown in black and Se in green. The ``zigzag'' direction is along the x-axis (a lattice vector) while the ``armchair'' is along the y-axis (b lattice vector).}
    \label{fig:fig1}
\end{figure}

With the introduction of the charged vacancy, the atomic structure around the defect is distorted as shown in Figure \ref{fig:fig1}; the nature of the physical distortion combined with the underlying symmetries of the material provides insight into the features of the defect-centered exciton as discussed below. With the presence of the defect, the in-plane lattice slightly distorts, with lattice vectors expanding by $\sim 0.3~\%$ along the zigzag and $\sim 1~\%$ along the armchair direction. Along the out-of-plane direction, for the pristine material, each selenium atom is paired with a nearest-neighbor germanium atom almost directly above or below it in the out-of-plane direction with a $\sim 2.5$ \AA{} bond-length. For the optimized defective structure, the most significant change is that the neighboring Ge atom above the missing Se atom (that would be paired with it) moves towards the vacancy, shifting by approximately $1.8$ \AA{}, about $2/3$ of the way towards the opposite face of the layer. This results in a distortion of the monolayer along the out-of-plane direction. In particular, the unpaired Ge atom now has four nearest-neighbor Ge atoms, each at a distance of $\sim 2.7$ \AA{}, instead of five nearest-neighbor Se atoms. As discussed below, this distortion leads to unexpected behavior of the exciton wavefunction along the out-of-plane direction.  

\begin{figure}[htbp]
    \centering
    \includegraphics[width=1.0\linewidth]{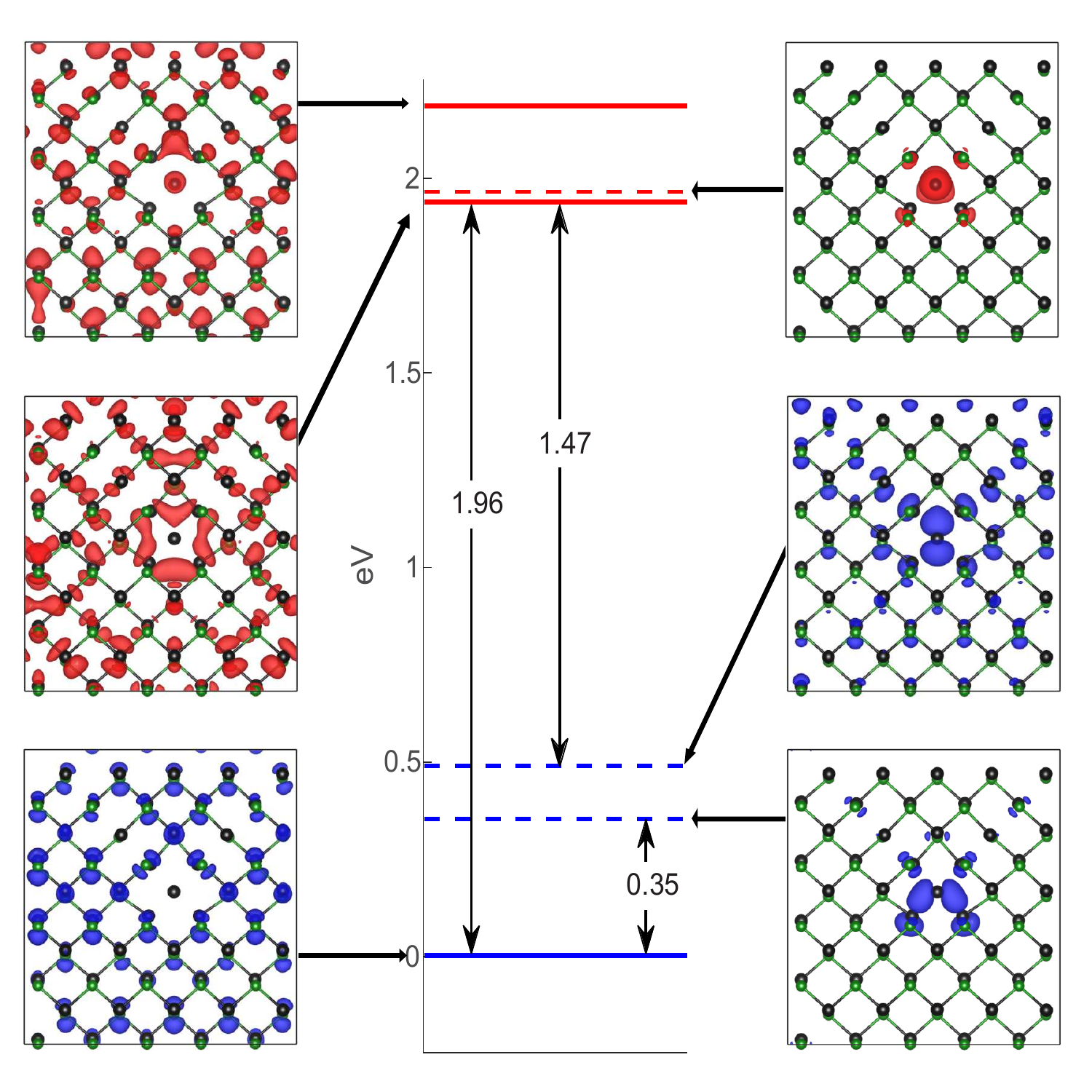}
    \caption{Schematic GW-predicted band diagram of monolayer GeSe containing the \vs{} point defect, with the pristine-like valence band maximum (VBM), conduction band minimum (CBM), and CBM+1 shown in solid lines and defect-centered states shown in dashed lines. The energy of the defect-centered states are taken as an average over the Brillouin Zone. Selected orbital densities at the \textbf{k} = $\Gamma$ point are also shown with an isosurface that encloses 33\% of the orbital density.}
    \label{fig:fig2}
\end{figure}

Figure \ref{fig:fig2} presents a schematic band-diagram for the defective semiconductor, showing the calculated defect state energy levels relative to the band edges of the pristine monolayer. We note that due to residual defect-defect interactions, there is some dispersion in the defect-centered energy level and thus Figure \ref{fig:fig2} presents the average energy associated with each defect-centered orbital. Previously, we determined that this average can accurately predict the converged defect-state energies for defective gallium nitride~\cite{lewis_quasiparticle_2017}. The \vs{} defect introduces two occupied defect-centered states within the band gap, at $0.35$ and $0.49$ eV above the pristine valence band edge, respectively. A third, unoccupied defect state is resonant with the conduction band edge. Because of the dispersive nature of the conduction band (0.34 eV bandwidth), the defect state can be either below or above this band depending on the position within the Brillouin zone; the average energy of the defect state is about $25$ meV above the CBM. The lowest energy gap between occupied and unoccupied states is from a defect-centered band within the gap to the unoccupied defect/pristine conduction band edge, with an energy difference of $1.47$ eV. This value is smaller than the predicted $1.9$ eV for the pristine monolayer, suggesting that the defect will result in low-energy optical transitions. 

\begin{figure}[htbp]
    \centering
    \includegraphics[width=1.0\linewidth]{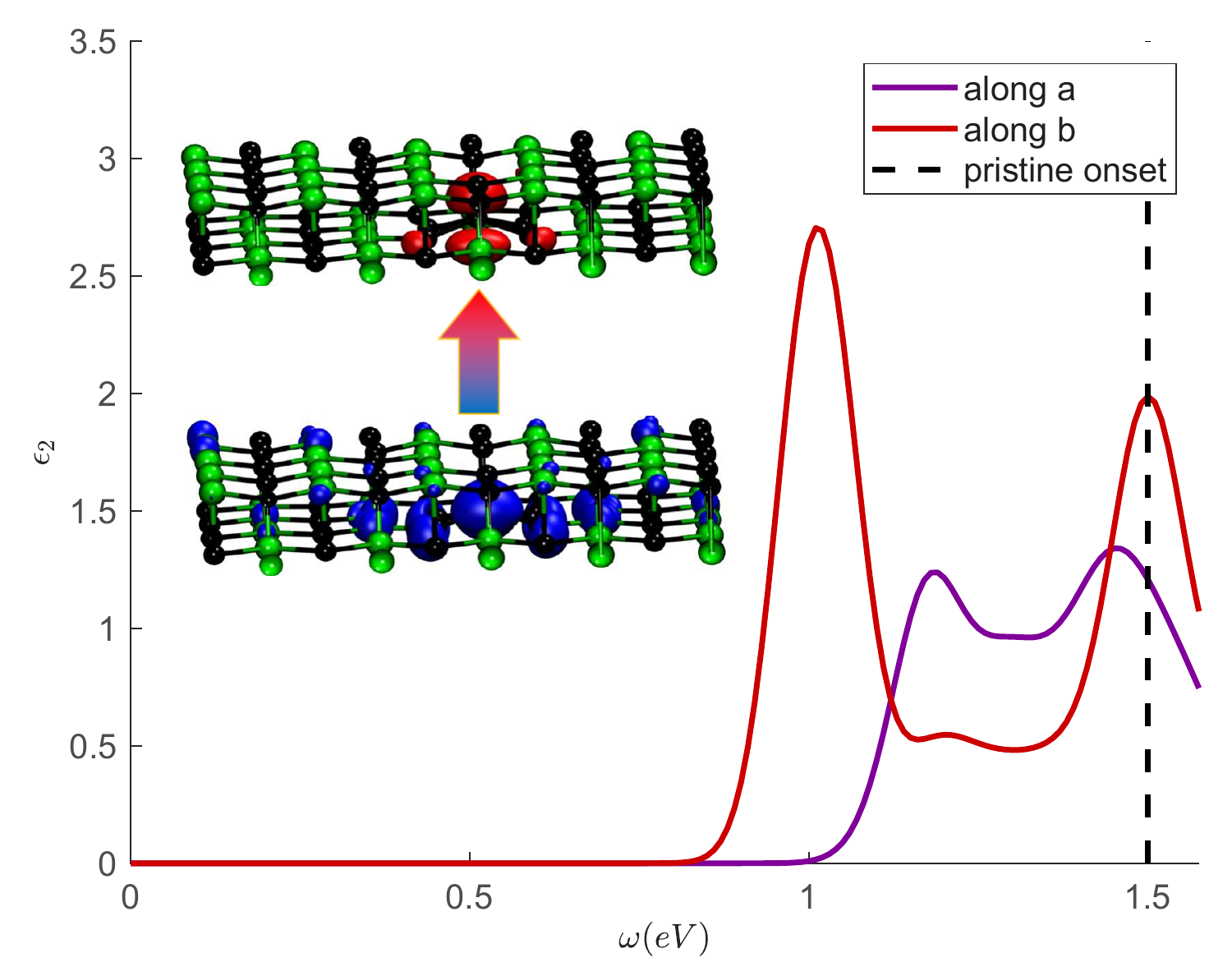}
    \caption{The imaginary component of the dielectric function with light polarized along the a (zigzag) and b (armchair) directions. The absorption onset for the pristine monolayer is indicated by the vertical dashed line. The inset shows the transition densities associated with the lowest energy exciton calculated as a weighted sum of BSE-predicted transitions, where the blue and red isosurfaces show the hole and electron components of the transition. }
    \label{fig:fig3}
\end{figure}

Figure \ref{fig:fig3} presents the predicted imaginary part of the dielectric function ($\epsilon_2$) along the a (zigzag) and b (armchair) directions. $\epsilon_2$ is anisotropic along these two axes, consistent with the anisotropic crystal structure. Such anisotropy has been predicted for the pristine monolayer GeSe as well~\cite{shi_anisotropic_2015,gomes_strongly_2016}. For the pristine monolayer, we predict an onset of absorption of $1.5$ eV, in agreement with previous studies~\cite{shi_anisotropic_2015}. The introduction of the defect results in lowering of the onset of absorption to $0.99$ eV, which is prominent in the absorption along the armchair direction. The inset of Figure \ref{fig:fig3} shows the transition associated with the lowest energy excited-state. Both valance and conduction orbitals are highly localized, suggesting that the exciton is composed of a defect-to-defect transition, with a lack of symmetry along the out-of-plane direction consistent with the structural distortions along that axis as described above. Interestingly, the exciton binding energy for this low-energy state (calculated as the energy difference between the free electron-hole pair and the bound exciton) remains unchanged with respect to the lowest-energy state of the pristine monolayer; both are predicted to be $0.3$ eV. For comparison, the exciton binding energy of bulk GeSe is predicted to be $<0.01 eV$~\cite{shi_anisotropic_2015}. This finding is consistent with the fact that the reduced screening in 2D materials results in a large increase in the exciton binding energy, but is not very sensitive to the degree of localization, particularly where the binding energy in the bulk is small~\cite{cudazzo_exciton_2016}.

Characterization of excitons in 2D materials, both experimentally and theoretically, is less straightforward than in bulk materials. The reduced screening often results in such significant increases in the exciton binding energy that the distinction between Wannier-Mott, charge transfer, and Frenkel excitons cannot be made reliably based on binding energy~\cite{cudazzo_exciton_2016}. Nevertheless, different classes of excitons have been predicted in pristine 2D materials based on visualization of the exciton wavefunction, including highly localized (Frenkel) excitons in SiC ~\cite{hsueh_excitonic_2011}, and relatively delocalized (Wannier-Mott) excitons in MoS\textsubscript{2}~\cite{qiu_optical_2013}, graphane~\cite{cudazzo_strong_2010, cudazzo_exciton_2016}, and pristine monolayer GeSe~\cite{shi_anisotropic_2015}.

To quantify the extent of the exciton in pristine and with-defect monolayer GeSe, we compute the electron-hole envelope correlation function (ECF), following previous studies~\cite{sharifzadeh_low-energy_2013, lewis_defect-induced_2019}. The ECF is calculated as
\begin{equation} \label{ecfint}
    \mathcal{F}(\textbf{r}) = \frac{\int |\Psi(\textbf{r}_{e}=\textbf{r}+\textbf{r}_{h},\textbf{r}_{h})|^{2}d^{3}\textbf{r}_{h}}{\int |\psi(\textbf{r}_{e})\psi(\textbf{r}_{h})|^{2}d^{3}\textbf{r}_{h}}, \\
\end{equation}

where $\textbf{r}_{e}$ and $\textbf{r}_{h}$ are the electron and hole coordinates, respectively, $\Psi(\textbf{r}_{e}=\textbf{r}+\textbf{r}_{h})$ is the two-particle wavefunction, and $\Psi(\textbf{r}_{e,h})$ are the non-interacting electron and hole single-particle wavefunctions. As discussed in Ref. \cite{lewis_defect-induced_2019}, $\mathcal{F}(\textbf{r})$, which is a function of distance between electron and hole, provides information about the exciton envelope wave function~\cite{yu_fundamentals_2010}, including an estimate of the Wannier-Mott Bohr radius. Additionally, the ECF allows identification of the charge transfer character of the exciton, by showing the relative distribution of electron and hole~\cite{sharifzadeh_low-energy_2013}.

Previously, we demonstrated that for a $1s$-type hydrogenic exciton envelope function such as a Wannier-Mott exciton, the ECF should take the form,
\begin{equation} \label{decay}
    \mathcal{F}^{Wannier}(r) = A_{0}~e^{(\frac{-2r}{a})},
\end{equation}
where $A_{0}$ is a normalization constant, r is the radial coordinate and a is the exciton Bohr radius. Here, we apply this analysis to the lowest energy exciton state of the pristine and with-defect GeSe. For the pristine GeSe, we predict a 2D Wannier-type exciton with a spatial extent (diameter) of $\sim 4$ nm, consistent with previous studies~\cite{shi_anisotropic_2015}. Fitting the radial distribution of the ECF,
\begin{equation} \label{radecf}
    \mathcal{F}(|r|) = (x^2 + y^2 +z^2)^\frac{1}{2},
\end{equation}
to Equation \ref{decay}, we predict a Bohr radius of $19.1$ \AA{}, as shown in the supplemental materials~\cite{cohen2020SM}.

\begin{figure}[htbp]
    \centering
    \includegraphics[width=1.0\linewidth]{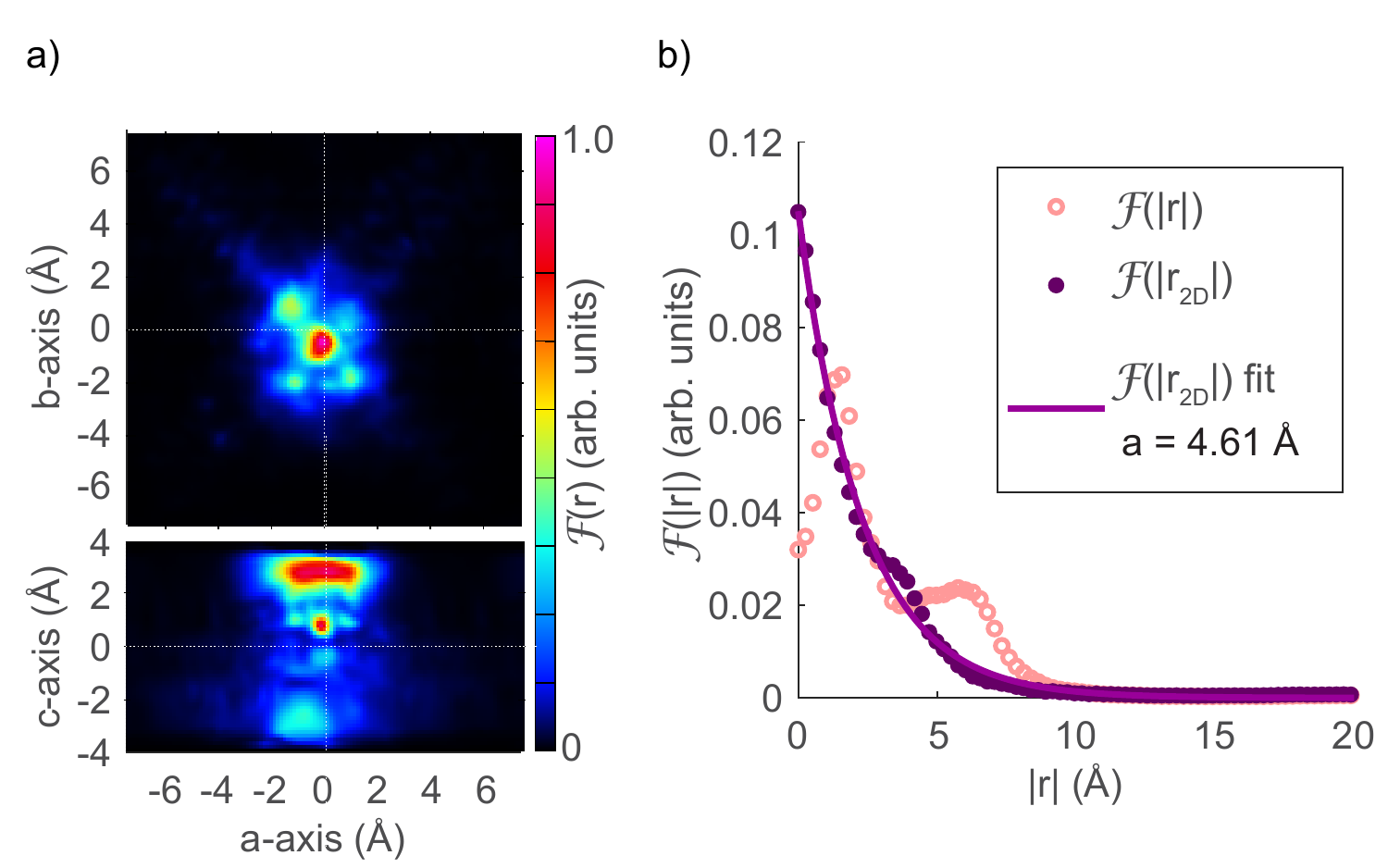}
    \caption{a) Two-dimensional ECF for with defect GeSe top view and side view. b) Radial ECF calculated in the a-b plane.}
    \label{fig:fig4}
\end{figure}

The distribution of the exciton is significantly altered by the presence of the defect. Figure \ref{fig:fig4}a shows the two-dimensional ECF for the lowest energy exciton within the a-b crystallographic plane, averaged along c (top panel) and the a-c crystallographic plane averaged along b (bottom panel). Within the a-b plane, $\mathcal{F}(\textbf{r})$ peaks near the origin (where electron and hole occupy the same space) and drops off quickly and symmetrically, consistent with a highly localized $1s$-type hydrogenic wavefunction. However, along the a-c plane, the exciton is highly directional, with a peak above the center of the monolayer. Electron and hole are preferentially separated such that along the c-axis, $\textbf{r}_e - \textbf{r}_h > 0 \sim 3$ \AA{} (i.e., the electron is more likely to be towards the top of the GeSe surface and the hole towards the bottom). This asymmetry reflects the different atomic environments on the two faces of the monolayer in the vicinity of the defect as discussed above. Interestingly, the asymmetry of the ECF suggests that the exciton has a dipole moment, pointed in the out-of-plane direction, which would not be present in the bulk. Such a phenomenon may also be present in other group IV-VI semiconductors with a similar hinge-like out-of-plane structure (such as GeS, SnSe and SnS).

Figure \ref{fig:fig4}b) presents the radial distribution of the ECF in defective GeSe~\cite{lewis_defect-induced_2019}. When all three dimensions are considered (see Equation \ref{radecf}), $\mathcal{F}(|r|)$ shows two features: one relatively sharp peak centered at $\sim 1.6$ \AA{} and a broader peak at $\sim 6$ \AA{} that does not show a simple exponential decay. The position and broadening of these peaks is due to the disorder introduced along the out-of-plane axis by the defect. In order to avoid the uncertainty in distribution associated with the symmetry breaking along the out-of-plane axis, we compute
\begin{equation} \label{rad2d}
    \mathcal{F}(|r|) = (x^2 + y^2)^\frac{1}{2}, 
\end{equation}
where the function is projected onto the 2D plane. The 2D function is now a decaying function consistent with a hydrogenic $1s$ state. Fitting this function, we find a Bohr radius of $4.6$\AA{} with an adjusted $R^{2} = 0.996$. In other words, when considered in the plane of the monolayer, the lowest energy exciton behaves as a highly localized Wannier-Mott exciton. However, when considered in the out-of-plane direction, the lowest energy exciton shows a charge-transfer character in that electron and hole are separated by $\sim 4$\AA{}.

Lastly, as noted above, the unoccupied defect-centered state is resonant with the conduction band of GeSe. Thus, the excitonic states are also a mixture of defect-like localized and Wannier-Mott-like excitons. In particular, the second excited state in the defective monolayer is more delocalized than the lowest energy transition; $\mathcal{F}(|r_{2D}|)$) can be fit to a hydrogenic wavefunction with Bohr radius $\sim 17.7$ \AA{}, comparable to the pristine GeSe, and the asymmetry in the out-of-plane direction is greatly reduced (see supplemental materials\cite{cohen2020SM}).

\section{Conclusions}
In summary, we applied many-body perturbation theory within the GW/BSE approximation to quantitatively describe excited-states in defective monolayer GeSe. We determined that the Se vacancy in the -2 charge state results in highly localized defect-centered electronic states and low-energy excitonic states, redshifting the onset of absorption by 0.5 eV. The lowest energy excitonic peak in the imaginary component of the dielectric function appears in the armchair direction, which is the direction in which the material can be most easily strained. Analysis of the lowest-energy exciton wavefunction suggests introduction of the defect strongly perturbs the pristine material's Wannier-Mott exciton, localizing it by a factor of four and introducing an asymmetry in the exciton distribution in the out-of-plane direction. These results underline the importance of considering defects in predicting the electronic and optical properties of monolayer materials. In the case of GeSe, defects with highly localized electronic states and unusual behavior in the out-of-plane direction could have potential for novel applications in 2D semiconductor devices such as photocatalysts.

\section{Acknowledgements}
The authors are grateful for funding from the U.S. Department of Energy (DOE), Office of Science, Basic Energy Sciences (BES) Early Career Program under Award No. DE-SC0018080. Additionally, we acknowledge grants of computer time from the National Energy Research Scientific Computing Center (NERSC), a DOE Office of Science User Facility supported by the Office of Science of the U.S. Department of Energy under Contract No. DE-AC02-05CH11231, the Extreme Science and Engineering Discovery Environment (XSEDE), which is supported by National Science Foundation Grant No. ACI-1548562, and Boston University Scientific Computing Center at the Massachusetts Green High-Performance Computing Center (MGHPCC).
\bibliography{References}

\end{document}